# "Numerical investigation of coherent and turbulent structures of light via nonlinear integral mappings"

A.Yu. Okulov

*Russian Academy of Sciences, 119991, Moscow, Russia*
*email: alexey.okulov@gmail.com ; tel/fax +7-495-382-4609*

The propagation of stable coherent entities of an electromagnetic field in nonlinear media with parameters varying in space can be described in the framework of iterations of nonlinear integral transformations. It is shown that for a set of geometries relevant to typical problems of nonlinear optics, numerical modeling by reducing to dynamical systems with discrete time and continuous spatial variables to iterates of local nonlinear Feigenbaum and Ikeda mappings and nonlocal diffusion-dispersion linear integral transforms is equivalent to partial differential equations of the Ginzburg-Landau type in a fairly wide range of parameters. Such nonlocal mappings , which are the products of matrix operators in the numerical implementation, turn out to be stable numerical-difference schemes, provide fast convergence and an adequate approximation of solutions. The realism of this approach allows one to take into account the effect of noise on nonlinear dynamics by superimposing a spatial noise specified in the form of a multimode random process at each iteration and selecting the stable wave configurations. The nonlinear wave formations described by this method include optical phase singularities, spatial solitons, and turbulent states with fast decay of correlations. The particular interest is in the periodic configurations of the electromagnetic field obtained by this numerical method that arise as a result of phase synchronization, such as optical lattices and self-organized vortex clusters.

---

# Численное моделирование когерентных и турбулентных структур излучения методом нелинейных интегральных отображений.

А.Ю.Окулов

*Российская академия наук*

E_mail :  alexey.okulov@gmail.com
url:  https://sites.google.com/site/okulovalexey/home

**Аннотация.**

Распространение устойчивых когерентных образований электромагнитного поля в нелинейных средах с меняющимися в пространстве параметрами может быть описано в рамках итераций нелинейных интегральных преобразований. Показано что для ряда актуальных геометрий задач нелинейной оптики численное моделирование путем сведения к динамическим системам с дискретным временем и непрерывными пространственными переменными, основанное на итерациях локальных нелинейных отображений Фейгенбаума и Икеды, а также нелокальных диффузионно-дисперсионных линейных интегральных преобразованиях, эквивалентно в довольно широком диапазоне параметров дифференциальным уравнениям в частных производных типа Гинзбурга-Ландау. Такие нелокальные отображения, представляющие собой при численной реализации произведения матричных операторов, оказываются устойчивыми численно-разностными схемами, обеспечивают быструю сходимость и адекватную аппроксимацию решений. Реалистичность данного подхода позволяет учитывать влияние шумов на нелинейную динамику путем наложения на расчетный массив чисел при каждой итерации пространственного шума, задаваемого в виде многомодового случайного процесса и производить отбор устойчивых волновых конфигураций. Нелинейные волновые образования, описываемые данным методом, включают оптические фазовые сингулярности, пространственные солитоны и турбулентные состояния с быстрым затуханием корреляций. Определенный интерес представляют полученные данным численным методом периодические конфигурации электромагнитного поля, возникающие в результате фазовой синхронизации, такие как оптические решетки и самоорганизованные вихревые кластеры.

*Дискретные отображения, интегральные преобразования, солитоны, вихри, фронты переключения, вихревые решетки, хаос, турбулентность.*

## Введение.

Спонтанное образование сложных пространственно-временных структур в распределенных нелинейных системах в присутствии шумов представляет интерес с точки зрения физики конденсированного состояния [1], нелинейной оптики и лазерной физики. Сверхпроводники II рода во внешнем магнитном поле демонстрируют образование гексагональных решеток незатухающих токов вокруг силовых линий магнитной индукции, проникающей в сверхпроводник [2]. Тонкий слой оптического материала с кубичной нелинейностью показателя преломления и зеркалом обратной связи формирует треугольные решетки интенсивности [3]. Широкоапертурный твердотельный лазер излучает сфазированные решетки оптических вихрей прямоугольной симметрии [4,5,6]. Подобные самоорганизующиеся структуры аккуратно описываются в рамках уравнений типа Гинзбурга-Ландау. Области устойчивости структур определяются путем расчета инкрементов неустойчивости (показателей Ляпунова) [7]. При численном моделировании необходимо обеспечить сходимость к решению, аппроксимирующему реальную ситуацию и проверить это решение на устойчивость [8]. Наиболее распространенные численные методы включают в себя неявные схемы Кранка-Николсона [9], метод расщепления по дисперсии (дифракции) и нелинейности с использованием быстрого преобразования Фурье [10], метод конечных элементов [11]. Во всех случаях ключевым моментом является идентификация численных артефактов, которые могут генерироваться самой разностной схемой как устойчивые пространственно-временные образования, но не иметь при этом никакого отношения, ни к решаемому нелинейному волновому уравнению, ни, тем более, к реальной физической ситуации, описываемой этим уравнением. В данной работе предлагается физически мотивированный подход к численному моделированию сложных нелинейных систем использующий произведения интегральных операторов диффузионного типа и точечных нелинейных преобразований, хорошо исследованных в теории динамического хаоса. Этот подход сочетает физическую наглядность математических структур и возможность универсальным образом описывать базовые нелинейные волновые образования в присутствии шумов.

**Моделирование стационарных режимов генерации.**

В ряде практически интересных случаев геометрия решаемой задачи позволяет существенно упростить численное моделирование и повысить уровень достоверности получаемых результатов. Одна из типичных геометрий в лазерной физике адекватно моделируется итерационным методом интегрального уравнения Фокса-Ли [12], где комплексные массивы $E_{n,m}$ эмулируют распределение амплитуды и фазы световой волны на зеркалах, а распространение излучения рассчитывается в рамках классической теории дифракции [13]. Данный метод естественным образом включает в разностную схему граничные условия, позволяет учесть качественно нелинейность усиливающей среды [14] и обладает высокой степенью устойчивости. В достаточно общей форме итерационный метод Фокса-Ли записывается как:

$$E_{n+1}(\vec{r}) = \int_{-\infty}^{\infty} K(\vec{r}-\vec{r}\,') f[E_n(\vec{r}\,')] d\vec{r}\,', \quad (1)$$

где $K(\vec{r} - \vec{r}\,')$ функция Грина линейного волнового уравнения, $E_{n+1}(\vec{r})$ - распределение комплексной амплитуды световой волны (электрическое поле в уравнениях Максвелла), записываемое в расчетной схеме как комплексный массив $E_{n,m}$, $f[E_n(\vec{r})]$ - нелинейная передаточная функция усиливающей среды, являющаяся решением уравнений Максвелла – Блоха в усиливающем элементе без учета дифракции [15]. Собственные функции и собственные значения интегрального уравнения (1) дают стационарные нелинейные моды $\widetilde{E}(\vec{r})$, собственные частоты $\mathrm{Im}\,\Gamma$ и декременты затухания $\mathrm{Re}\,\Gamma$. Следующим последовательным приближением к реальной нелинейной динамике, наблюдаемой экспериментально, является учет релаксации активной среды [16], описываемый временем жизни атомов на верхнем уровне $T_1$ (продольной релаксации):

$$\frac{N_{n+1} - N_n}{\Delta t} = +\frac{N_0 - N_n}{T_1} - \sigma N_n |E_n|^2, \qquad (2)$$

где $N_n$ - число возбужденных атомов в единице объема, $N_0 / T_1$ - скорость накачки атомов (скорость перевода атомов на верхний уровень), $\sigma$ - сечение вынужденного излучения [13,16]. Вместе с передаточной функцией усиливающей среды (на примере двухуровневого лазерного усилителя [6,7]):

$$E_{n+1} = f[E_n(\vec{r})]] = R E_n \cdot \exp[i k n_0 L_c + i k n_2 |E_n|^2 L_{amp}] \cdot \exp[\sigma N_n L_{amp}], \qquad (3)$$

комбинации из уравнений вида (1-3) дают полное описание пространственно-временной нелинейной динамики лазеров на диэлектрических кристаллах и полупроводниковых лазеров с т.н. «вертикальным резонатором» (рис.1), где $k = 2\pi / \lambda$, $R$ - коэффициент отражения выходного зеркала, $n_0$ -линейный показатель преломления, $n_2$ - нелинейный показатель преломления, $L_c$ -длина резонатора, $L_{amp}$ -толщина усиливающего элемента.

Другой типичной нелинейностью является квадратичная восприимчивость $\chi_2$, используемая для генерации оптических гармоник и параметрического преобразования частоты излучения. В простейшем случае, когда при условии фазового синхронизма гармоник происходит эффективное преобразование частоты, есть возможность включить в интегральное уравнение Фокса-Ли в качестве нелинейной передаточной функции одномерное отображение следующего вида [17]:

$$E_{n+1} = f[E_n(\vec{r})] = g(E_n)[1 - \tanh(g(E_n))] \quad . \quad (4)$$

Физическим образом отображения (4) это является внутрирезонаторная генерация второй оптической гармоники в кольцевом лазере. Циклические проходы излучения через усиливающую среду $g(E_n)$ чередуются с преобразованием излучения в нелинейном кристалле с квадратичной восприимчивостью $\chi_2$ [18]. Итерации отображений с квадратичным максимумом в виде простейшей логистической параболы $E_{n+1} = G E_n [1 - E_n^2]$ и (4) $E_{n+1} = G E_n [1 - \tanh(E_n)]$ подчиняются универсальным закономерностям Фейгенбаума [19]. Бифуркационные диаграммы таких отображений имеют универсальную структуру (рис.1).

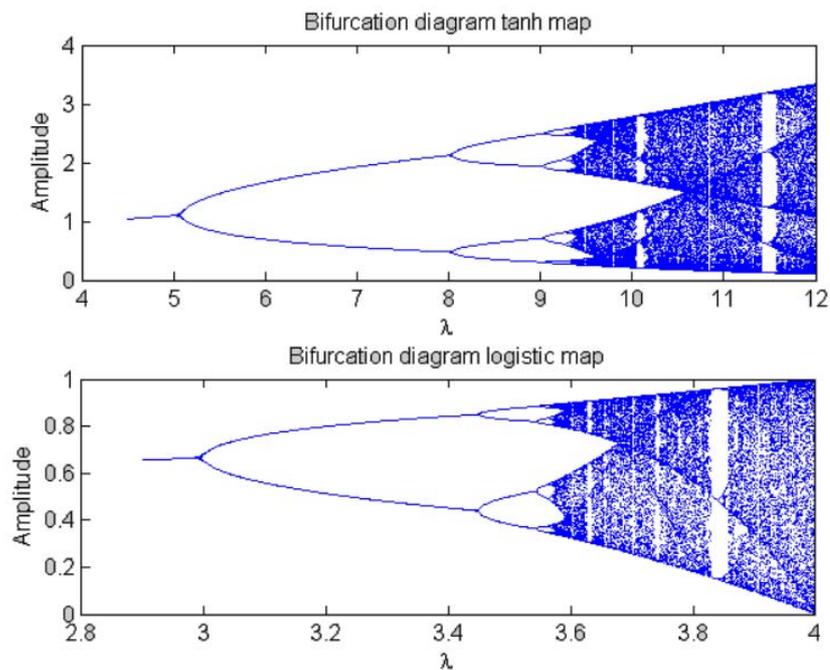

Рис.1

Бифуркационные диаграммы а) отображения (4) и б) логистического отображения демонстрируют скейлинг с универсальными числами $\delta_F = 4.66$ и $\alpha_F = 2.502$ вблизи точек перехода к хаосу $G_{chaos} \cong 9.4$ (а) и $G_{chaos} \cong 3.99...$ (б).

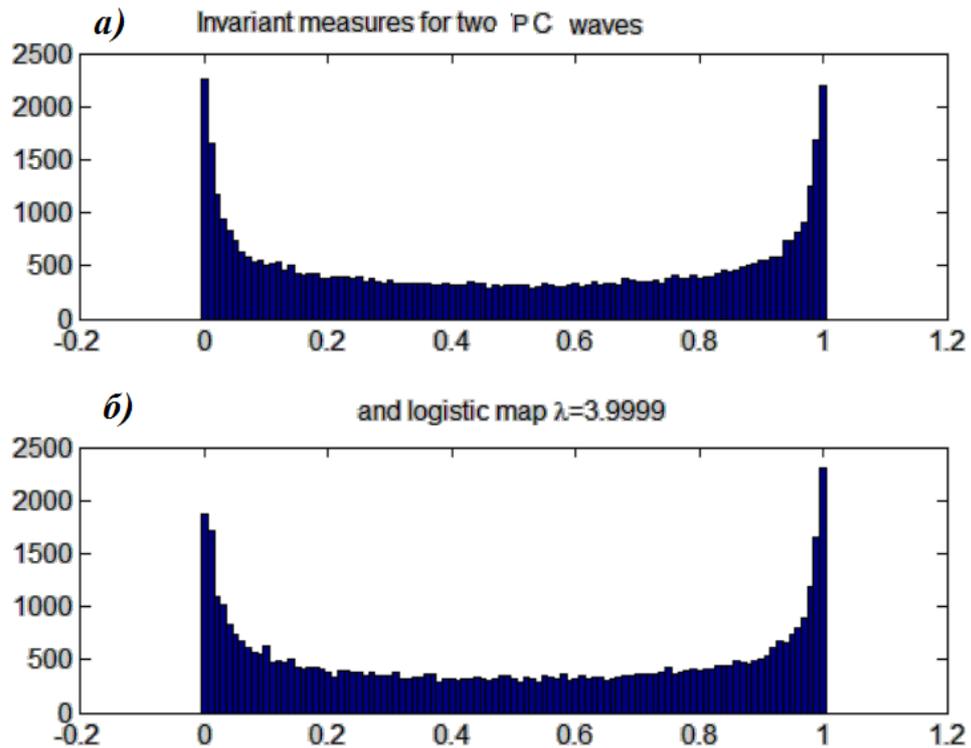

Рис.2

а) Гистограмма интерференции двух фазово-сопряженных волн со случайными фазами [20] и б) гистограмма в точке перехода к хаосу $G_{chaos} = 3.99$ логистического отображения.

Значения бифуркаций критического параметра (в данном случае это ненасыщенный коэффициент усиления $G$) сгущаются вблизи точек перехода к хаосу асимптотически как геометрическая прогрессия:

$$\delta_F = \lim_{M \to \infty} \frac{\lambda_M - \lambda_{M-1}}{\lambda_{M-1} - \lambda_{M-2}} \to 4.6692... \qquad (5)$$

В точках перехода в хаосу, например при значении коэффициента усиления $G_{chaos} = 9.4$ для отображения (4), итерации поля $E_n$ дают хаотический набор чисел, статистика которого в точности описывается $\beta$-распределением [19]. Примечательно, что в точности аналогичное $\beta$-распределение вероятности:

$$P(I) = \frac{1}{\pi I \sqrt{1-I}}, \, I \approx I_0[1 + \cos(\varphi_1 - \varphi_2)], (\varphi_1 - \varphi_2) \subset (0, \pi) \quad , \qquad (6)$$

получается и при интерференции двух волн со случайными фазами [20], где $I$ интенсивность интерферирующих волн с разностью фаз $\varphi_1 - \varphi_2$, равномерно распределенной в интервале $(0, \pi)$.

Итерационный метод Фокса – Ли с такой стационарной моделью усиливающей среды на основе уравнений (2) и (3) продемонстрировал сходимость к решениям в виде гауссовых пусков и периодических решеток с самовоспроизведением полей за счет эффекта Тальбота и позволил проверить эти решения на устойчивость [21]. Неожиданным оказалось то обстоятельство, что геометрия оптического резонатора с тонким слоем усиливающей среды, накачиваемым вдоль оси генерации излучением лазера накачки, выбранная с точки зрения максимальной простоты математической модели (1), [15,17,18,21], оказалась чрезвычайно интересной с точки зрения экспериментаторов [22], поскольку лазер с активной средой в виде тонкого диска имеет ряд существенных преимуществ с точки зрения контроля тепловых режимов и расходимости.

**Моделирование нестационарных режимов генерации.**

Реальная динамика твердотельных и полупроводниковых лазеров характеризуется релаксационными колебаниями, которые возникают вследствие существенного количественного различия скорости релаксации инверсной населенности (время продольной релаксации $T_1$ или время жизни атома на верхнем уровне) и скорости релаксации поля в резонаторе (время жизни фотона в резонаторе $\tau_c$). В простейшей модели одномодового одночастотного лазера:

$$\frac{dE}{dt} + \frac{E}{\tau_c} = \frac{\sigma N\, E}{2} + \delta E \quad ; \quad \frac{dN}{dt} + \frac{N - N_0 + \delta N}{T_1} = -\sigma N |E|^2 \,, \qquad (7)$$

линеаризованный анализ устойчивости стационарного режима генерации [13] дает следующее выражение для периода затухающих колебаний, возникающих при отклонении динамической системы (6) от равновесия:

$$\tau_{REL} = \sqrt{\frac{T_1 \tau_C}{G-1}} \quad , \quad G = \sigma N_0 \, ; \quad \nu_{REL} = 1/\tau_{REL} \quad . \qquad (8)$$

Отклонения от равновесия в системе уравнений (6) возникают вследствие шумов, т.е. флуктуаций инверсной населенности $\delta N$ и поля $\delta E$ в резонаторе. Спектр мощности излучения $I(\omega)$ имеет в результате флуктуаций характерную форму с максимумом на частоте релаксационных колебаний $\nu_{REL}$ [4]. Аналогичным образом записывается система уравнений с дискретным временем на основе нелокального отображения Фокса-Ли и уравнения для релаксирующей инверсной населенности (1-3):

$$E_{n+1}(\vec{r}) = \int_{-\infty}^{\infty} K(\vec{r}-\vec{r}\,')f[E_n(\vec{r}\,')]d\vec{r}\,' + \delta E_n(\vec{r});$$

$$f(E_n(\vec{r})) = RE_n(\vec{r}) \cdot \exp[ikn_0 L_c + ikn_2|E_n(\vec{r})|^2 L_{abs}] \cdot \exp[\sigma N_n(\vec{r}) L_{amp}];$$

$$\frac{N_{n+1}(\vec{r}) - N_n(\vec{r})}{\Delta t} = +\frac{N_0(\vec{r}) + \delta N(\vec{r}) - N_n(\vec{r})}{T_1} - \sigma N_n(\vec{r})|E_n(\vec{r})|^2; \quad (9)$$

$$K(\vec{r}-\vec{r}\,') = \frac{ik}{2\pi L_c}\exp\left[\frac{ik|\vec{r}-\vec{r}\,'|^2}{L_c}\right];$$

Эта система имеет качественно близкий спектр мощности $I(\omega)$ с релаксационным максимумом на частоте релаксационных колебаний $\nu_{REL}$ [16], который наблюдался в ряде экспериментов с твердотельными чип-лазерами с диодной оптической накачкой [4]. Кроме того, вследствие нелинейного взаимодействия поперечных типов колебаний в спектре мощности возникают дополнительные пики, обусловленные возбуждением коллективных колебаний образовавшейся оптической решетки. Пучности образовавшейся оптической решетки ведут себя подобно атомам в кристаллической решетке, коллективные колебания которой имитируют возбуждения «акустических» и «оптических» фононов [5]. Дополнительный «оптический» резонанс $\nu_{opt}$ в спектре мощности расположен дальше от несущей частоты $\nu_0$, чем «акустический» $\nu_{ac}$ и самый медленный из них релаксационный резонанс $|\nu_{REL} - \nu_0| < |\nu_{ac} - \nu_0| < |\nu_{opt} - \nu_0|$ [16]. На рис.3 изображены фазовые портреты в переменных $(\text{Re}\,E_n, \text{Im}\,E_n)$, $(I(t), N(t))$ при малом превышении над порогом генерации $G = \sigma N_0 L_{amp} \approx 1.2$, при существенном различии времен жизни фотона в резонаторе $\tau_C \approx 10^{-8}$ sec и продольной релаксации $T_1 \approx 10^{-4}$ sec, а также числе Френеля $N_f \approx N_p^2 \approx 100$ [21], соответствующем возбуждению квадратной решетки интенсивности на выходном зеркале из 9x9 ($N_p \times N_p$) пятен.

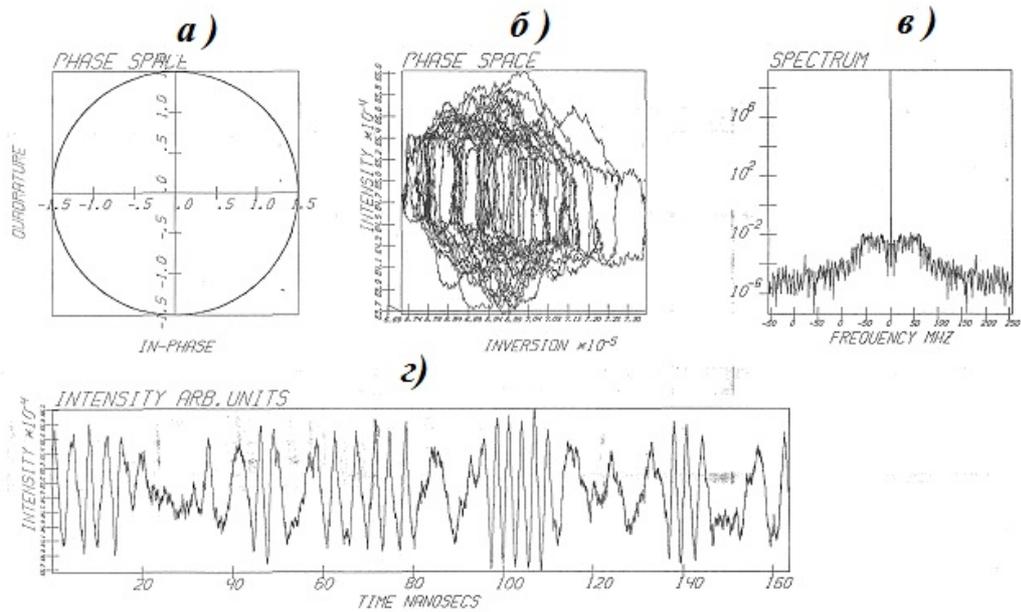

Рис.3

а) Фазовый портрет динамической системы (8) в переменных $(\operatorname{Re} E_n, \operatorname{Im} E_n)$ или (in-phase, quadrature) при отсутствии шумов, б) Фазовый портрет системы (8) в переменных $(I(t), N(t))$ со случайными шумами $\delta N_n(\vec{r})$, $\delta E_n(\vec{r})$, добавляемыми на каждой итерации (проходе излучения через резонатор), в) спектр мощности $I(\omega)$ системы (8) при малом превышении порога генерации, г) временная реализация интенсивности излучения системы (8) на оси резонатора $I_n(t) = |E_n(\vec{r})|^2$.

Особенностями динамического режима, полученного итерациями нелокального отображения (8), является относительно слабый пик на частоте релаксационных колебаний $\nu_{REL} = \sqrt{\dfrac{G-1}{T_1 \tau_C}} \approx 50 Mhz$ (рис.3в), обусловленный малым превышением коэффициента усиления $G$ над порогом генерации. Если при отсутствии шумов (рис.3а) фазовый портрет системы (8) близок к когерентному состоянию [23], т.е. круговой фазовой траектории в переменных $(\operatorname{Re} E_n, \operatorname{Im} E_n)$ слегка размытой за счет соотношения неопределенностей, то при наличии шумов, обусловленных спонтанным распадом верхнего уровня с временем жизни $T_1$, флуктуациями инверсной населенности $\delta N_n(\vec{r})$ вследствие пульсаций оптической накачки и флуктуациями поля в резонаторе $\delta E_n(\vec{r})$, включающими спонтанное излучение, фазовая траектория в переменных $(I(t), N(t))$ (рис.3б) все время выталкивается из фокуса бесшумовой $(\delta E_n(\vec{r}), \delta N_n(\vec{r}) = 0)$ динамической системы (6) $\delta$-образными случайными толчками. Соответственно фазовая траектория в переменных $(I(t), N(t))$ в этом реалистичном режиме оказывается составленной из большого количества «броуновских» изломов, обусловленных случайными силами $\delta N_n(\vec{r})$, $\delta E_n(\vec{r})$, возмущающими обе динамические переменные: поле в резонаторе $E(t)$ и инверсную населенность $N(t)$. Очевидно, что ответ на вполне логичный вопрос о том, являются ли фазовые траектории системы (8) фракталом или одним из классических шумовых процессов [23] требует чрезвычайно аккуратного численного моделирования, которое могло бы различить классическую шумовую траекторию от автомодельной (self-similar) траектории фрактала, воспроизводящей себя при последовательном увеличении разрешения расчетной сетки. Определенный прогресс в этом направлении был достигнут благодаря работам Грассбергера и Прокацци [24], которые разработали остроумный алгоритм восстановления фазового пространства динамической системы по реализациям временного ряда всего лишь одной динамических переменных системы и вычисления фрактальной размерности аттрактора в том случае, когда система находится в режиме динамического хаоса. Однако такие расчеты довольно объемны и требуют определенных затрат машинного времени и вычислительных ресурсов даже для простейших систем [25].

**Вихревые решетки.**

Пространственная структура излучения, содержащаяся в двумерных комплексных массивах $E_n(\vec{r})$, выводилась в виде двух распределений интенсивности $I_n(\vec{r}) = E_n(\vec{r}) \cdot E_n^*(\vec{r})$ и оптической фазы $\arg E_n(\vec{r})$. Размерность массивов $E_n(\vec{r})$ и $N_n(\vec{r})$ варьировалась в диапазоне от $128 \times 128$ до $512 \times 512$, хотя в некоторых тестах, проведенных для идентификации возможных численных артефактов, размерность массивов достигала $1024 \times 1024$. В отдельных тестах использовалась максимально возможная для используемого компьютера с процессором Pentium Dual Core и оперативной памятью 2 Гб размерность $2048 \times 2048$. В отличие от оригинальной вычислительной работы [5] и натурных экспериментов начатых в [4], где главное внимание было уделено распределению интенсивности $I_n(\vec{r})$, в данной работе главное внимание уделялось фазовой структуре излучения, которая содержит информацию о моменте импульса излучения. Фазовые сингулярности, т.е. точки, где модуль $|E_n(\vec{r})|$ а также интенсивность $I_n(\vec{r})$ обращаются в 0, окружены областями, где фаза светового поля $\arg E_n(\vec{r})$ меняется на величину $\ell\, 2\pi$ при обходе сингулярности по замкнутому контуру, где $\ell$ - топологический заряд сингулярности, а $\ell\,\hbar$ - орбитальный момент импульса фотонов в таком оптическом вихре. В большинстве реализаций, полученных путем итерирования изначально шумового поля $E_0(\vec{r})$ при числе Френеля $N_f \approx 80 \div 1000$, устанавливался квазистационарный режим генерации с квадратной решеткой пятен интенсивности (рис. 4), но наиболее интересная информация содержалась в фазовом распределении, которое явно указывало на существование решетки фазовых сингулярностей (рис. 5).

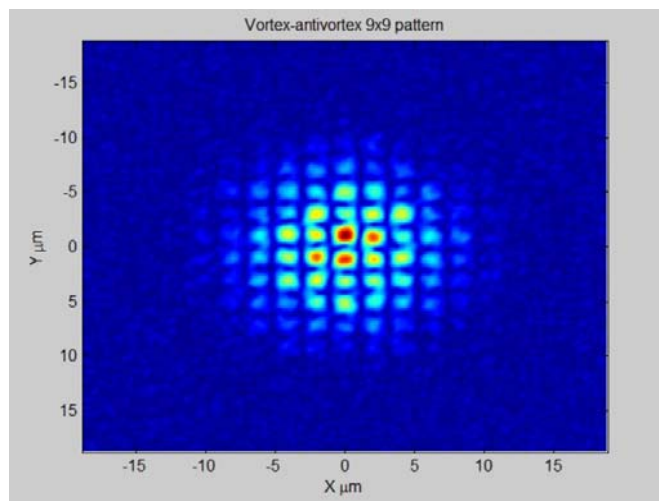

Рис 4.

Квадратная решетка из 9x9 оптических вихрей с противоположными циркуляциями на выходном зеркале твердотельного чип-лазера с параметрами резонатора из [4] и [6].

Изменение цвета от синего до красного при обходе вокруг этих особых точек означает изменение аргумента комплексного поля $E_n(\vec{r})$ от 0 до $\ell\, 2\pi$. В подавляющем большинстве реализаций оказалось, что топологические заряды (орбитальные квантовые числа $\ell$) расположены в «шахматном порядке». Элементарная ячейка вихрей состоит из четырех сингулярностей, локализованных по вершинам квадрата, причем соседние сингулярности имеют противоположные заряды $\pm\ell$, а сингулярности на диагонали имеют одинаковые по знаку топологические заряды. Эффективное «поле скоростей», рассчитываемое по преобразованию Маделунга $\vec{V}_\perp \approx grad[\arg E_n(\vec{r})]$ в поперечной плоскости или проекция на плоскость $\vec{r} = [x, y]$ импульса электромагнитного поля $\vec{P}_\perp = \varepsilon_0 c^2 [\vec{E} \times \vec{B}]_\perp$ состоит из решетки ячеек с противоположными циркуляциями скорости вокруг особых точек с нулевой интенсивностью (рис. 5). Полученные решетки вихрь – антивихрь отличаются от вихревых решеток в сверхпроводниках II рода, где вихри сонаправлены и образуются вокруг линий магнитной индукции, проникающей в тонкую пленку сверхпроводника и наводящей вокруг проникшего магнитного поля незатухающие круговые токи [2].

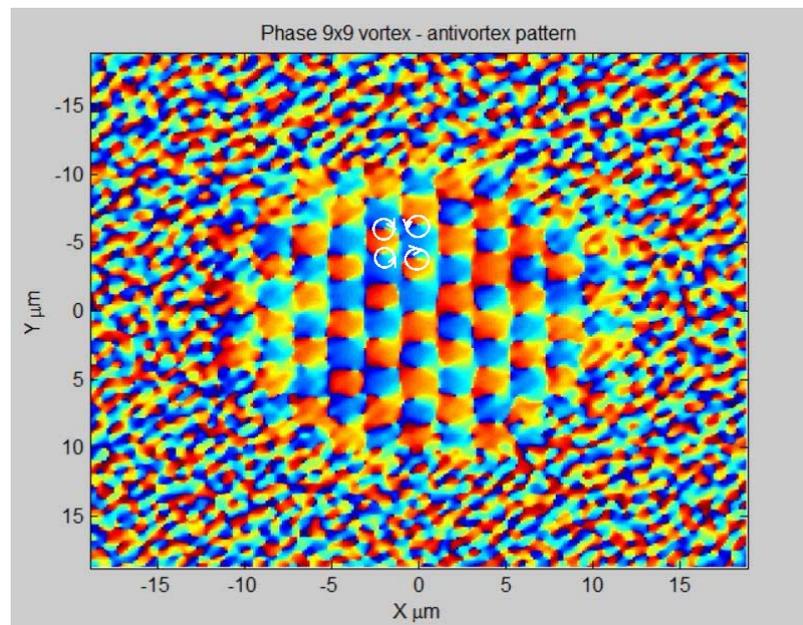

Рис.5

Фаза квадратной решетки из 9x9 оптических вихрей с противоположными циркуляциями. Изменение фазы дается изменением цвета от синего ($\varphi = 0$) до красного ($\varphi = 2\pi$). Мелкомасштабный фон из хаотически расположенных вихрей является многомодовым случайным процессом $\delta E_n(\vec{r})$, добавляемым к полю $E_n(\vec{r})$ на каждой итерации $n \to n+1$ (проходе через резонатор).

**Моделирование локализованных структур.**

Нелинейное взаимодействие поперечных мод оптического резонатора является базовым механизмом образования локализованных структур.

Возникающие в ближнем поле многомодового лазера пространственные солитоны [7], вихревые оптические солитоны («темные солитоны») и солитонные кластеры являются результатом развития модуляционной неустойчивости нелинейного уравнения Шредингера [27]. Численное моделирование этих волновых образований требует тщательной настройки численного кода и контроля процесса расчета для идентификации численных артефактов. В отличие от «светлых» солитонов [26], представляющих собой области максимумов световой волны с плавным профилем интенсивности вблизи максимума, вихревые солитоны и вихревые решетки состоят из оптических сингулярностей, где градиенты поля велики и это обстоятельство накладывает жесткие требования на разрешение расчетной сетки. Это обстоятельство существенно для светлых солитонов, вокруг которых зачастую возникает интерференционная картинка, обусловленная интерференцией солитона с фоном (background) . В этом случае также возникают резкие градиенты световой волны, требующие измельчения расчетной сетки. На рис.7 представлен результат расчета локализованного солитонного возбуждения возникшего на фоне гладкого фона. В нелокальном отображении (8) использовались два расчетных массива $E_n(\vec{r})$ комплексной огибающей световой волны и массив комплексного коэффициента усиления $N_n(\vec{r})$, $n_n(\vec{r})$, включающий распределение инверсной населенности $N_n(\vec{r})$ и показателя преломления $n_n(\vec{r})$ на однородной расчетной сетке $512 \times 512$ точек.

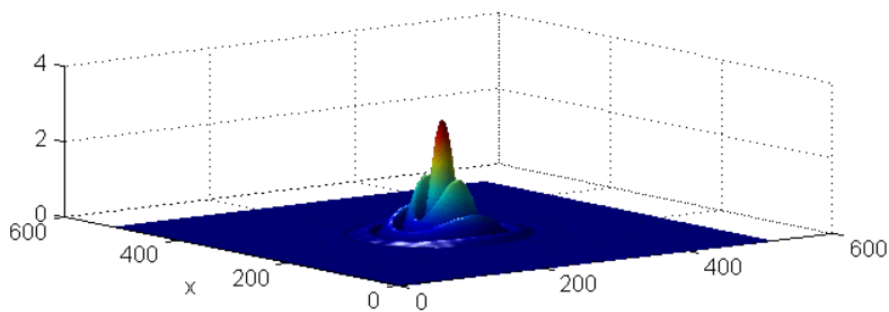

Рис.6

Локализованное возбуждение на выходном зеркале $E_n(\vec{r})$ как результат синхронизации поперечных мод устойчивого резонатора. Осцилляции вокруг резкого центрального пика обусловлены интерференцией пространственного солитона с фоном.

**Заключение.**

Использование нелокальных отображений для расчета нестационарных нелинейных волновых процессов позволяет существенно упростить численное моделирование сложных оптических систем. Физически наглядное построение вычислительной процедуры из чередующихся интегралов свертки и точечных преобразований [6, 14-16, 21] дает возможность контролировать устойчивость и сходимость алгоритма путем фильтрации нежелательных пространственных гармоник и получать реалистичные численные решения, совпадающие в ряде точно решаемых случаев с аналитическими результатами.

**Литература.**